\documentclass[twocolumn, amssymb, jcp, superscriptaddress,showpacs, nobibnotes]{revtex4-1}

\usepackage[dvips]{graphicx}

\begin{document}

\title{Photodissociation and radiative association of HeH$^+$ in the metastable triplet state}

\author{J. Loreau}
\affiliation{ITAMP, Harvard-Smithsonian Center for Astrophysics, Cambridge, Massachusetts 02138, USA}
\affiliation{Service de Chimie Quantique et Photophysique, Universit\'e Libre de Bruxelles (ULB) CP 160/09, 1050 Brussels, Belgium}
\author{S. Vranckx}
\affiliation{Service de Chimie Quantique et Photophysique, Universit\'e Libre de Bruxelles (ULB) CP 160/09, 1050 Brussels, Belgium}
\affiliation{Laboratoire de Chimie Physique, UMR8000, Universit\'e de Paris-Sud, Orsay 91405, France.}
\author{M. Desouter-Lecomte}
\affiliation{Laboratoire de Chimie Physique, UMR8000, Universit\'e de Paris-Sud, Orsay 91405, France.}
\author{N. Vaeck}
\affiliation{Service de Chimie Quantique et Photophysique, Universit\'e Libre de Bruxelles (ULB) CP 160/09, 1050 Brussels, Belgium}
\author{A. Dalgarno}
\affiliation{ITAMP, Harvard-Smithsonian Center for Astrophysics, Cambridge, Massachusetts 02138, USA}

\begin{abstract}
We investigate the photodissociation of the metastable triplet state of HeH$^+$ as well as its formation through the inverse process, radiative association. In models of astrophysical plasmas, HeH$^+$ is assumed to be present only in the ground state, and the influence of the triplet state has not been explored. It may be formed by radiative association during collisions between a proton and metastable helium, which are present in significant concentrations in nebulae. 
The triplet state can also be formed by association of He$^+$ and H, although this process is less likely to occur.
We compute the cross sections and rate coefficients corresponding to the photodissociation of the triplet state by UV photons from a central star using a wave packet method. We show that the photodissociation cross sections depend strongly on the initial vibrational state and that the effects of excited electronic states and non-adiabatic couplings cannot be neglected. We then calculate the cross section and rate coefficient for the radiative association of HeH$^+$ in the metastable triplet state.

\end{abstract}

\maketitle

\section{Introduction}

The helium hydride ion HeH$^+$ is one of the most elementary molecular ions and the first to form in the early universe \cite{Lepp2002}, through the direct radiative association process \cite{Roberge1982}
\begin{equation}\label{eq_rad_stab_X}
\mathrm{He}(1s^2) + \mathrm{H}^+ \longrightarrow \mathrm{HeH}^+(X\ ^1\Sigma^+) + h\nu \ ,
\end{equation}
which occurs in the ground state of the molecular ion. 
In addition, HeH$^+$ can also be formed by a spontaneous radiative transition from the vibrational continuum of the excited $A\ ^1\Sigma^+$ state to a discrete vibrational level of the ground state \cite{Zygelman1990,Kraemer1995}:
\begin{equation}\label{eq_rad_stab_A}
\mathrm{He}^+(1s) + \mathrm{H}(1s) 
\longrightarrow \mathrm{HeH}^+(X\ ^1\Sigma^+) + h\nu
\end{equation}

These mechanisms of production of HeH$^+$ were investigated along with destruction processes in order to estimate the abundance of the molecular ion in the early universe as well as in various astrophysical environments. 
In particular, HeH$^+$ was predicted to be observable in planetary and gaseous nebulae such as NGC 7027 \cite{Black1978,Flower1979,Roberge1982,Cecchi-Pestellini1993}. However, despite searches for vibrational or rotational lines, no emission from HeH$^+$ has been detected from these objects so far, although it has been observed in laboratory plasmas for many years \cite{Hogness1925,Bernath1982}. 
It was shown that the observation of the $J=1-0$ rotational line of HeH$^+$ is hindered by the near-coincidence of this line with a CH rotational line that has a greater intensity \cite{Liu1997}.
Other possibilities to detect HeH$^+$ include the potential presence of the ion in helium-rich white dwarfs \cite{Gaur1992,Harris2004}, metal-poor stars \cite{Engel2005}, or in supernovae \cite{Miller1992}, but the molecular ion has so far eluded observation.

The formation of HeH$^+$ is mainly due to the radiative association between He and H$^+$ or between He$^+$ and H (processes (\ref{eq_rad_stab_X}) and (\ref{eq_rad_stab_A})). While it has always been supposed that HeH$^+$ is formed in its ground $X\ ^1\Sigma^+$ state, one should also consider the possible role of the first metastable triplet state, $a\ ^3\Sigma^+$. This state, which correlates asymptotically to He$^+(1s$) + H($1s$), can indeed be populated and will not decay by collisions if the plasma density is low. As its radiative decay to the ground state is spin-forbidden, the state is metastable with a lifetime of 150~s \cite{Loreau2010c}. 
Moreover, in experimental studies on the dissociative recombination of HeH$^+$ \cite{Yousif1994, Strasser2000}, it was suggested that the $a\ ^3\Sigma^+$ state might be present in the ion beam and be responsible for a part of the cross section.

In this work, we study the main mechanisms controlling the abundance of HeH$^+$ in the $a\ ^3\Sigma^+$ state. The first reaction of interest is the photodissociation process, which destroys the molecular ion:
\begin{eqnarray}\label{eq_photodissociation}
\mathrm{HeH}^+ (a\ ^3\Sigma^+) + h\nu \longrightarrow (\mathrm{HeH}^+)^*
&& \longrightarrow \mathrm{He}(1snl\ ^3L) + \mathrm{H}^+ \nonumber \\
&& \longrightarrow \mathrm{He}^+(1s) + \mathrm{H}(nl) \nonumber \\
\end{eqnarray}
The photodissociation can occur following excitation to $^3\Sigma^+$ or $^3\Pi$ electronic states and leads to atomic fragments He + H$^+$ or He$^+$ + H. We investigate this reaction with a time-dependent method based on the propagation of a wave packet on the excited electronic states coupled by non-adiabatic interactions that gives access to the contribution of each excited state to the cross section. 

Based on the results on the photodissociation process, we examine the radiative association of HeH$^+$ in the $a\ ^3\Sigma^+$ state, which is one of the main mechanism of formation of this state. It occurs through a spontaneous radiative transition from the vibrational continuum of the excited $b\ ^3\Sigma^+$ state to a discrete vibrational level of the metastable $a\ ^3\Sigma^+$ state:
\begin{equation}\label{eq_radiative_association}
\mathrm{He}(1s2s\ ^3S) + \mathrm{H}^+ 
\longrightarrow \mathrm{HeH}^+ (a\ ^3\Sigma^+) + h\nu
\end{equation}
The $b\ ^3\Sigma^+$ molecular state corresponds to collisions between H$^+$ and He in its metastable $1s2s\, ^3S$ state, which has a lifetime of about 8000 s \cite{Drake1971}. Since metastable helium is easily produced in an ion source, the reaction (\ref{eq_radiative_association}) could lead to the presence of HeH$^+$ in the $a\ ^3\Sigma^+$ state in the source. Since the $b\ ^3\Sigma^+$ state has a large potential well (0.73~eV), it could also be produced \cite{Vranckx2012}, and its radiative decay would result in populating the $a\ ^3\Sigma^+$ state as well, a process that was investigated by Chibisov {\it et al.} \cite{Chibisov1996}. 
Metastable helium is also present in various astrophysical environments \cite{Scherb1968,Indriolo2009}, and in particular in planetary nebulae \cite{Drake1972,Clegg1987}, mostly due to the recombination of He$^+$ with free electrons. The radiative association process (\ref{eq_radiative_association}) might therefore lead to the presence in space of HeH$^+$ in its triplet state.
HeH$^+$ can also be formed in the $a\ ^3\Sigma^+$ state by a transition from the continuum to a bound ro-vibrational state of the same potential energy curve. However, the rate coefficient for this process is extremely small \cite{Kraemer1995}.

We start in Sec. \ref{section_theory} by summarizing previous results of the molecular structure of HeH$^+$ as well as the theory of the photodissociation and radiative association processes. We present the photodissociation and radiative association cross sections and rate coefficients in Sec. \ref{section_results}, and we discuss possible applications of the $a\ ^3\Sigma^+$ state.

\section{Theoretical Methods}\label{section_theory}

\subsection{Molecular data}

\begin{figure}
\includegraphics[width=.5\textwidth]{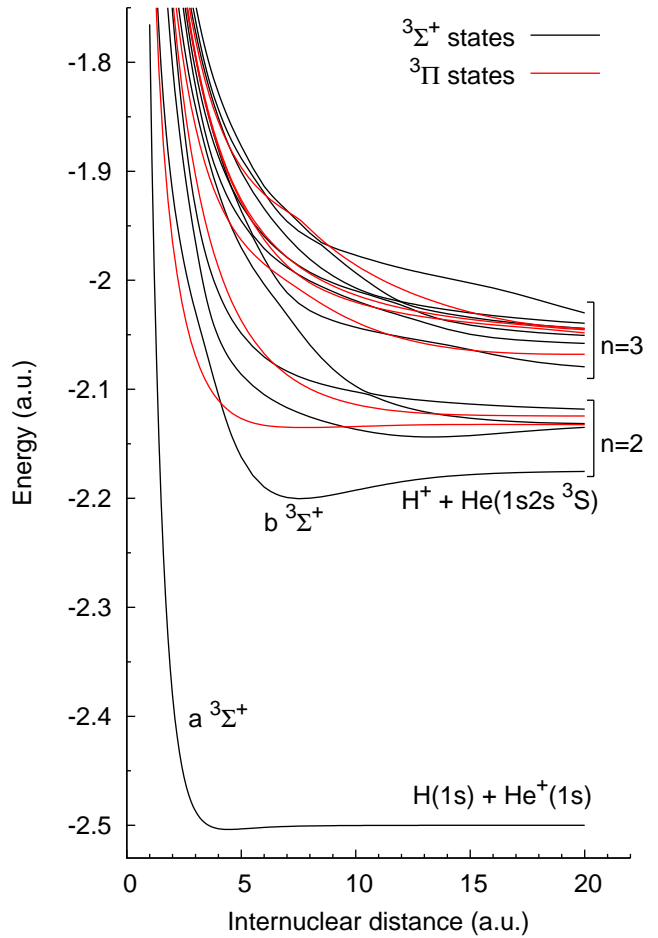}
  \caption{Adiabatic PEC of the $n=1-3$ triplet states of HeH$^+$ from Ref. \cite{Loreau2010a}. Black, $^3\Sigma^+$ states; Red, $^3\Pi$ states. The $a\ ^3\Sigma^+$ and $b\ ^3\Sigma^+$ states are shown with their dissociation limits.}
  \label{fig_elec_states}
\end{figure}

We consider here all the $^3\Sigma^+$ and $^3\Pi$ electronic states dissociating into atomic states with $n=1-3$, where $n$ is the largest principal quantum number of the atomic fragments. There is only one $n=1$ $^3\Sigma^+$ state, while there are 4 $^3\Sigma^+$ and 2 $^3\Pi$ $n=2$ states, as well as 6 $^3\Sigma^+$ and 4 $^3\Pi$ $n=3$ states. This makes a total of 11 $^3\Sigma^+$ states and 6 $^3\Pi$ states. The adiabatic potential energy curves (PEC) for these triplet states are shown in Fig. \ref{fig_elec_states}. 
The low-energy electronic spectrum of the molecular ion consists of states dissociating either into H + He$^+$ or into H$^+$ + He. For example, the lowest triplet state ($a\ ^3\Sigma^+$) dissociates into H($1s$) + He$^+(1s)$, while the first excited state ($b\ ^3\Sigma^+$) dissociates into H$^+$ + He($1s2s\, ^3S$). 
We adopted the molecular data presented in Ref. \cite{Loreau2010a}. The potential energy curves, non-adiabatic couplings and dipole transition moments were calculated at the complete active space self-consistent field (CASSCF) and configuration interaction (CI) levels. 
Since the excited electronic states undergo avoided crossings (see Fig. \ref{fig_elec_states}), the radial non-adiabatic couplings must be taken into account. 
Due to the large number of such couplings, we do not show them here and refer the reader to Refs. \cite{Loreau2010a,Loreau2011c} where they are discussed.
The matrix of the radial non-adiabatic couplings in the basis of the adiabatic electronic functions $\{\zeta_m\}$, $F_{mm^{\prime}}=\langle\zeta_m \vert \partial_{R} \vert \zeta_{m^{\prime}} \rangle$, can be used to build the diabatic representation \cite{Smith1969}. The adiabatic-to-diabatic transformation matrix $\mathbb{D}(R)$ is found by solving the differential matrix equation
\begin{equation}\label{diab_representation}
\partial_R\mathbb{D}(R)+\mathbb{F}(R)\cdot \mathbb{D}(R)=0
\end{equation}
As in previous studies involving the excited states of HeH$^+$ \cite{Sodoga2009,Loreau2010b}, we have only retained the couplings between adjacent states, $F_{m,m\pm 1}$. Eq. (\ref{diab_representation}) was solved numerically by imposing the initial condition $\mathbb{D}(\infty)=\mathbb{I}$. The diabatic potential energy curves are the diagonal elements of the matrix $\mathbb{U}^\mathrm{d}=\mathbb{D}^{-1}\cdot \mathbb{U\cdot D}$, where $\mathbb{U}$ is the matrix of $H^{\mathrm{el}}$ in the adiabatic representation, and the off-diagonal elements of $\mathbb{U}^\mathrm{d}$ are the diabatic couplings. 
On the other hand, we neglected the non-adiabatic rotational couplings, as their effect is expected to be negligible \cite{Loreau2011b}.

\subsection{Photodissociation cross sections and rates}

The photodissociation cross section from a given rovibrational state $v,J$ is computed with a time-dependent wave packet method that has been described in detail in Refs. \cite{Sodoga2009,Loreau2010b}, so we only briefly outline its main features.
The first step is to calculate the rovibrational energies $E_{vJ}$ and wave functions $\psi_{vJ}(R)$ of the $a\ ^3\Sigma^+$ state, which is achieved using a B-spline method. The potential supports six vibrational states for $J=0$, as was reported in Ref. \cite{Loreau2010c}. The ground vibrational state is bound by 664.8 cm$^{-1}$ while the $v=5$ state is bound by less than 2 cm$^{-1}$, and there is a total of 51 bound rovibrational states.

The wave packet at $t=0$, $\Phi_{vJ}^{f}(R,t=0)$ is defined on each excited state $f$ as the product of the wave function of the rovibrational state by the dipole matrix element $\mu_{if}(R)=\langle \zeta_i \vert d \vert \zeta_f \rangle$  between the initial state $i$ and the final state $f$:
\begin{equation} \label{photodiss_initial}
\Phi_{vJ}^{f}(R,t=0) = \mu_{if}(R) \psi_{vJ}(R) \ .
\end{equation}
In this case, the initial state $i$ is the $a\ ^3\Sigma^+$ state while the final state $f$ can be any excited $^3\Sigma^+$ or $^3\Pi$ state. The transition to $\Sigma$ states occurs for a parallel orientation of the field with respect to the molecular axis, while the transition to $\Pi$ states occurs for a perpendicular orientation.
The wave packet subsequently evolves on the coupled potential energy curves, and the time-propagation is performed with the split-operator method in the diabatic representation \cite{Feit1982,Alvarellos1988}. 
The total photodissociation cross section can then be computed from the Fourier transform of the autocorrelation function $C(t)= \sum_f	 \langle \Phi_{vJ}^{f}(R,0) \vert \Phi_{vJ}^{f}(R,t) \rangle$ \cite{Heller1978a}.
The propagated wave packet therefore contains the information about the cross section for all energies. 

However, in this work we are not interested in the total cross section but rather in the partial photodissociation cross sections for each of the excited electronic states. We wish to understand which states form the dominant contribution to the cross section, and the partial cross sections are also necessary if one wants to compute the radiative association cross section (see below).
The computation of the partial cross sections is realized using a method based on the Fourier transform of the wave packet at a point $R_{\infty}$ located in the asymptotic region \cite{Balint-Kurti1990}. The partial cross section for each final excited state $f$ is given by
\begin{equation}\label{photodiss_part_CS_R}
\sigma_{vJ}^{f} (E_{\text{ph}}) = \frac{4\pi^2 \alpha a_0^2 k_f}{\mu} E_{\text{ph}}  \big\vert  A_{vJ}^{f}  \big\vert ^2 \ ,
\end{equation}
where $\mu$ is the reduced mass of the system, $E_{\text{ph}}=h\nu$ is the photon energy, $k_f=\sqrt{2\mu(E_{vJ}+E_{\text{ph}}-E^{f}_{\mathrm{as}})}$ is the wave number in the electronic state $f$ with an asymptotic energy $E^{f}_{\mathrm{as}}$, and
\begin{equation}\label{eq_fourier_transform}
A_{vJ}^{f} = \frac{1}{\sqrt{2\pi}} \int \Phi_{vJ}^{f}(R_{\infty},t) e^{i(E_{vJ}+E_{\text{ph}})t} dt \ .
\end{equation}

The calculations were performed using a spatial grid of $2^{13}$ points from 0.5 to 100 au. The point at which the Fourier transform (\ref{eq_fourier_transform}) is evaluated was chosen as $R_\infty=75$ au. To avoid unphysical reflexions of the wave packet at the end of the grid, an absorbing potential starting at $R=80$ au was introduced. The time step used in the propagation was 1 au, and tests with shorter steps were performed to assess the convergence of the cross sections. For the highest vibrational level, $v=5$, we extended the grid to 150 au and took $R_\infty=100$ au. This is necessary as the wave function for this weakly bound level extends to large internuclear distances.

The rate coefficient for the photodissociation following the absorption of a photon emitted by a central star of radius $R_{\star}$ with blackbody temperature $T_\star$ at a radial distance $R$ is given by \cite{Roberge1982}
\begin{equation}\label{eq_rate_photo}
k(T_\star)=\frac{2\pi}{h^3c^2} \bigg(\frac{R_\star}{R} \bigg)^2 \int \frac{E_{\text{ph}}^2}{e^{E_{\text{ph}}/k_BT_\star}-1} \sigma(E_{\text{ph}}) \ dE_{\text{ph}}\ ,
\end{equation}
We have chosen the value $(R_\star/R)^2=10^{-13}$, as suggested in Ref. \cite{Roberge1982} .

\subsection{Radiative association}

The cross section for the radiative association process can be obtained from the photodissociation cross section using the detailed balance principle. 
The radiative association cross section for the formation of the $a\ ^3\Sigma^+$ state into a level $v,J$ along the potential of any excited state can be calculated as \cite{Puy2007}		
\begin{equation}\label{eq_rad_stab_cs}
\sigma^{\text{a}}_{vJ}(E_k)=\frac{E_{\text{ph}}^2}{\mu c^2 E_k}\, \sigma^{\text{d}}_{vJ}(E_{\text{ph}})\ .
\end{equation}
In this equation, $\sigma^{\text{d}}_{vJ}(E_{\text{ph}})$ is the partial cross section corresponding to the photodissociation of the $v,J$ level of the $a\ ^3\Sigma^+$ state via an excited state as a function of the photon energy, while $\sigma^{\text{a}}_{vJ}(E_k)$ is the cross section for radiative association into the $v,J$ level as a function of the relative kinetic energy of the incident particles. It is related to the photon energy by
\begin{equation}\label{eq_rad_assoc_energy}
E_k= E_{\text{ph}} - \Delta E \ ,
\end{equation}
where $\Delta E$ is the difference between the dissociation energy of the electronic state along which the collision takes place and the energy of the rovibrational level of the $a\ ^3\Sigma^+$ state in which the association is realized.
As the association can occur into any bound rovibrational level, the cross section for radiative association must be summed over all possible values of $v$ and $J$.
It should be noted that, by construction, the radiative association cross section from a particular electronic state calculated using Eq. (\ref{eq_rad_stab_cs}) includes non-adiabatic effects.

The radiative association rate constant is calculated assuming a Maxwell-Boltzmann distribution of incident particles,
\begin{equation}\label{eq_rad_stab_rate}
k(T) = \Big(\frac{2}{k_BT}\Big)^{3/2} \frac{1}{\sqrt{\pi\mu}} \int_0^\infty E_k e^{-E_k/k_BT} \sigma(E_k) \ dE_k \ .
\end{equation}

\section{Results and discussion}\label{section_results}

\subsection{Photodissociation}

\begin{figure}
\includegraphics[angle=-90,width=.5\textwidth]{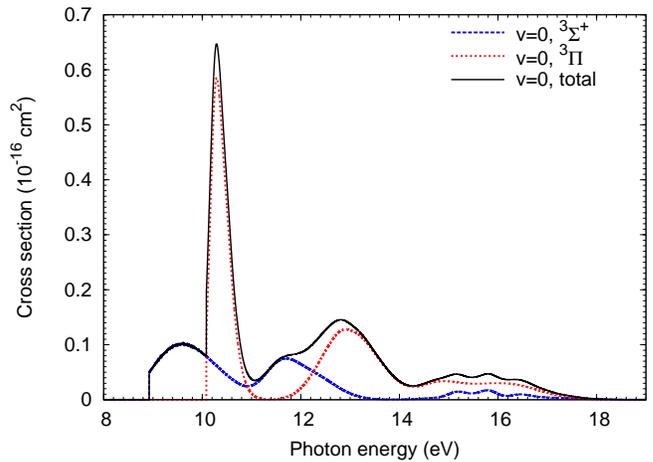}
  \caption{Photodissociation cross section for the initial state $v=0, J=0$. Blue dashed lines: contribution of the $^3\Sigma^+$ states. Red dotted lines: contribution of the $^3\Pi$ states. Black full line: total cross section.}
  \label{fig_photo_cs_v0}
\end{figure}

The photodissociation cross section for the initial rovibrational level $v=0, J=0$ is presented in Fig. \ref{fig_photo_cs_v0}. The total cross section, as well as the contributions of the $^3\Sigma^+$ and $^3\Pi$ states, are shown.
In the case of dissociation through $^3\Sigma^+$ states, the cross section is composed of two peaks centered around 9.5 eV and 11.6 eV. The first peak is due to the $b$ state and the second to the three remaining $n=2$ states. The small structure in the range $15-17$ eV is due to the $n=3$ states. 
We note that the cross section starts abruptly at 8.92 eV, which corresponds to the energy difference between the initial state ({\it i.e.}, the $v=0$ level of the $a\ ^3\Sigma^+$ state) and the dissociation energy of the $b\ ^3\Sigma^+$ state. 
The $b\ ^3\Sigma^+$ state has a deep potential well of about 0.73 eV, as can be seen from Fig. \ref{fig_elec_states}, with an equilibrium position $R_e=7.75 a_0$ \cite{Loreau2010a}. A component of the initial wave packet (\ref{photodiss_initial}) will have an energy below the dissociation energy of the $b\ ^3\Sigma^+$ state, and this component will oscillate in the potential well. Therefore, it never reaches the asymptotic region and makes no contribution to the cross section of the $b\ ^3\Sigma^+$ state. On the other hand, the component of the wave packet with sufficient energy will travel toward the asymptotic region, which explains the threshold behavior of the cross section.
As in the case of the photodissociation from the ground $X\, ^1\Sigma^+$ state of HeH$^+$ \cite{Sodoga2009}, we observed a strong influence of the non-adiabatic couplings on the photodissociation cross sections, due to the large number of avoided crossings affecting the excited states PEC (see Fig. \ref{fig_elec_states}).

The cross section for dissociation via $^3\Pi$ states is dominated by a narrow peak centered at 10.3 eV and a wider peak around 13 eV, both due to the two $n=2$ states. The shape of the cross section can be explained using the reflection principle, which states that the cross section reflects the spatial distribution of the wave function of the initial state, and that its width is proportional to the steepness of the potential energy curves in the excitation region \cite{Schinke1993}. As can be seen in Fig. \ref{fig_elec_states}, in the excitation region the PEC of the lowest $^3\Pi$ state is almost flat, leading to a very narrow cross section, while that of the second $^3\Pi$ state is steeper, resulting in a wider cross section.
The diffuse structure in the cross section between 14 and 17 eV is due to the $n=3$ states. As these states interact strongly through non-adiabatic couplings, the cross section is much less structured than for the two $n=2$ states.

\begin{figure}
\includegraphics[angle=-90,width=.5\textwidth]{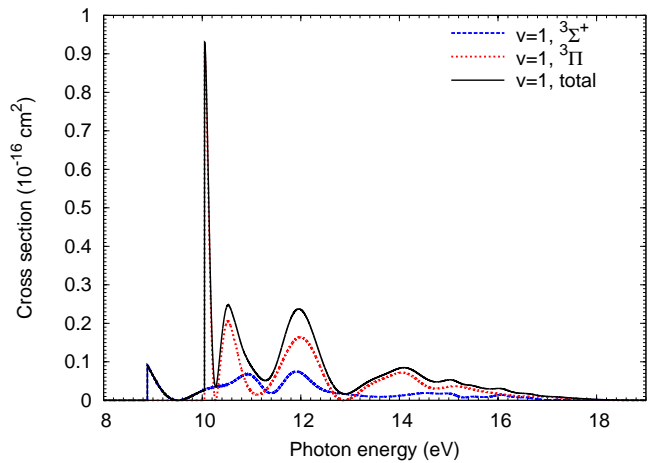}
  \caption{Photodissociation cross section for the initial state $v=1, J=0$. Blue dashed lines: contribution of the $^3\Sigma^+$ states. Red dotted lines: contribution of the $^3\Pi$ states. Black full line: total cross section.}
  \label{fig_photo_cs_v1}
\end{figure}

The photodissociation cross section is strongly dependent on the initial vibrational level. This is illustrated in Fig. \ref{fig_photo_cs_v1}, which shows the cross section for the initial state $v=1, J=0$. It has a more complicated structure than for the $v=0$ case, reflecting the oscillatory behavior of the initial wave function, but we still observe the dominance of the $^3\Pi$ states as well as a threshold behavior at an energy corresponding to the dissociation energy of the $b\ ^3\Sigma^+$ state. We reach similar conclusions for the higher vibrational levels $v\geq 2$.
On the other hand, the cross section is almost independent of $J$ and is insensitive to rotational excitation as the initial wave function depends only weakly on $J$.

The photodissociation rate coefficient, given by Eq. (\ref{eq_rate_photo}), is presented in Tab. \ref{table_photo_rate} as a function of the star temperature $T_\star$ and for three different value of the matter temperature $T_{\text{m}}$. The matter temperature $T_{\text{m}}$ governs the rovibrational distribution of the initial state of HeH$^+$. Assuming equilibrium, the population of each state is given by 
\begin{equation}\label{}
p_{vJ}(T) = (2J+1)\exp(-E_{vJ}/k_BT_{\text{m}}) / Z(T_{\text{m}})
\end{equation}
where $k_B$ is the the Boltzmann constant and $Z(T_{\text{m}})$ is the partition function.
The photodissociation cross sections and rate coefficients must take this dependence into account. 
We give in Tab. \ref{table_photo_rate} the contribution of the $b\ ^3\Sigma^+$ state to the rate coefficient (denoted by $k_b$), as well as the total contribution of the $^3\Sigma^+$ and $^3\Pi$ states ($k_{\Sigma}$ and $k_{\Pi}$, respectively) and the total rate ($k_{\text{tot}}$). We observe that even at low temperature, it is not sufficient to consider only the $b\ ^3\Sigma^+$ state to obtain an accurate value of the photodissociation rate coefficient. Other excited electronic states, and the $^3\Pi$ states in particular, must also be taken into account. For example, at $T_\star=10^4$~K the contribution of the $b\ ^3\Sigma^+$ state to the total rate coefficient is less than 50\% while at $T_\star=5\times 10^4$~K it does not exceed 20\%.
This is in stark contrast to the photodissociation from the ground state of HeH$^+$, in which case the rate is dominated by dissociation into the first excited state \cite{Loreau_thesis}.

\begin{table*}
\begin{ruledtabular}
  \caption{Photodissociation rate coefficients (s$^{-1}$). $T_{\text{m}}$(K) is the matter temperature and $T_{\star}$(K) is the temperature of the star. $k_b$ denotes the contribution of the $b\ ^3\Sigma^+$ state to the rate coefficient, $k_{\Sigma}$ is the contribution of all the $^3\Sigma^+$ states, $k_{\Pi}$ is the contribution of the $^3\Pi$ states, and $k_{\text{tot}}=k_{\Sigma}+k_{\Pi}$ is the total rate.}
  \label{table_photo_rate}
  \begin{tabular}{cccccc} 
$T_{\text{m}}$(K)	& $T_{\star}$(K)	& $k_b$		& $k_{\Sigma}$			& $k_{\Pi}$			& $k_{\text{tot}}$			\\ \hline
50			& $1\times 10^4$& $1.69\times 10^{-10}$	& $1.84\times 10^{-10}$	& $1.61\times 10^{-10}$	& $3.44\times 10^{-10}$		\\ 
			& $2\times 10^4$& $4.60\times 10^{-8}$	& $6.02\times 10^{-8}$	& $8.05\times 10^{-8}$	& $1.41\times 10^{-7}$		\\ 
			& $5\times 10^4$& $1.54\times 10^{-6}$	& $2.62\times 10^{-6}$	& $4.59\times 10^{-6}$	& $7.21\times 10^{-6}$		\\ 
			& $1\times 10^5$& $6.35\times 10^{-6}$	& $1.22\times 10^{-5}$	& $2.32\times 10^{-5}$	& $3.54\times 10^{-5}$		\\ 
			& $2\times 10^5$& $1.76\times 10^{-5}$	& $3.57\times 10^{-5}$	& $7.06\times 10^{-5}$	& $1.06\times 10^{-4}$		\\  
			& $5\times 10^5$& $5.29\times 10^{-5}$	& $1.11\times 10^{-4}$	& $2.23\times 10^{-4}$	& $3.34\times 10^{-4}$		\\ \hline
500			& $1\times 10^4$& $1.11\times 10^{-10}$	& $1.31\times 10^{-10}$	& $1.43\times 10^{-10}$	& $2.74\times 10^{-10}$		\\ 
			& $2\times 10^4$& $3.00\times 10^{-8}$	& $4.67\times 10^{-8}$	& $7.34\times 10^{-8}$	& $1.20\times 10^{-7}$		\\ 
			& $5\times 10^4$& $1.01\times 10^{-6}$	& $2.20\times 10^{-6}$	& $4.18\times 10^{-6}$	& $6.38\times 10^{-6}$		\\ 
			& $1\times 10^5$& $4.19\times 10^{-6}$	& $1.05\times 10^{-5}$	& $2.10\times 10^{-5}$	& $3.15\times 10^{-5}$		\\ 
			& $2\times 10^5$& $1.16\times 10^{-5}$	& $3.11\times 10^{-5}$	& $6.38\times 10^{-5}$	& $9.48\times 10^{-5}$		\\  
			& $5\times 10^5$& $3.50\times 10^{-5}$	& $9.68\times 10^{-5}$	& $2.01\times 10^{-4}$	& $2.98\times 10^{-4}$		\\ \hline
5000			& $1\times 10^4$& $9.93\times 10^{-11}$	& $1.22\times 10^{-10}$	& $1.41\times 10^{-10}$	& $2.63\times 10^{-10}$		\\ 
			& $2\times 10^4$& $2.69\times 10^{-8}$	& $4.44\times 10^{-8}$	& $7.25\times 10^{-8}$	& $1.17\times 10^{-7}$		\\ 
			& $5\times 10^4$& $9.08\times 10^{-7}$	& $2.12\times 10^{-6}$	& $4.10\times 10^{-6}$	& $6.22\times 10^{-6}$		\\ 
			& $1\times 10^5$& $3.77\times 10^{-6}$	& $1.01\times 10^{-5}$	& $2.05\times 10^{-5}$	& $3.06\times 10^{-5}$		\\ 
			& $2\times 10^5$& $1.05\times 10^{-5}$	& $3.01\times 10^{-5}$	& $6.21\times 10^{-5}$	& $9.22\times 10^{-5}$		\\  
			& $5\times 10^5$& $3.15\times 10^{-5}$	& $9.39\times 10^{-5}$	& $1.96\times 10^{-4}$	& $2.90\times 10^{-4}$		\\ 
\end{tabular}
\end{ruledtabular}
\end{table*}

\subsection{Radiative association}

\begin{figure}
\includegraphics[angle=-90,width=.5\textwidth]{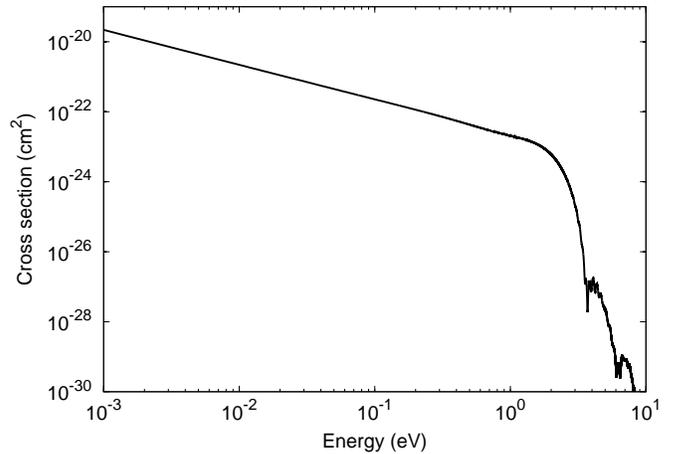}
  \caption{Radiative association cross section for the radiative association along the $b\ ^3\Sigma^+$ state, process (\ref{eq_radiative_association}).}
  \label{fig_rad_stab_cs_bstate}
\end{figure}

The radiative association cross section along the $b\ ^3\Sigma^+$ state, calculated using Eq. (\ref{eq_rad_stab_cs}), is presented in Fig. \ref{fig_rad_stab_cs_bstate} as a function of the collision energy. The cross section is dominated by transitions to the levels with $v=0-2$, and it is several orders of magnitude larger than the cross section for radiative association along the $a\ ^3\Sigma^+$ state that was studied Kraemer {\it et al.} \cite{Kraemer1995}. In that case, the association takes place via a radiative transition from the continuum to a bound level of the same potential, leading to a small cross section. The cross section shown in Fig. \ref{fig_rad_stab_cs_bstate} decreases sharply for energies over 2.77~eV, which corresponds to the threshold energy above which the transition probability from the $b\ ^3\Sigma^+$ state to the $a\ ^3\Sigma^+$ state becomes strongly suppressed. 
It should be noted that the cross section presented in Fig. \ref{fig_rad_stab_cs_bstate} is only the non-resonant part of the cross section and that the contribution from shape resonances due to the centrifugal potential, as well as resonances caused by the non-adiabatic interactions, should also be included. 
Unfortunately, the accurate determination of these resonances requires impractically long propagation times \cite{Vranckx2012}.
Taking shape resonances into account results in an enhancement of the radiative association rate coefficient and can be evaluated in a separate calculation using the Breit-Wigner theory \cite{Breit1936,Bain1972}. 
The contribution of resonances on the radiative association rate can be quite large in cold environments, but it decreases quickly with increasing temperature \cite{Gustafsson2012}. In addition, since the $a\ ^3\Sigma^+$ state has a small binding energy and supports only 51 bound rovibrational states, we can reasonably expect the number of shape resonances to be small and their contribution to the rate coefficient to be negligible except at very low temperature, similarly to what was recently shown for the radiative association of LiHe$^+$ \cite{Augustovicova2012}. 

\begin{figure}
\includegraphics[angle=-90,width=.5\textwidth]{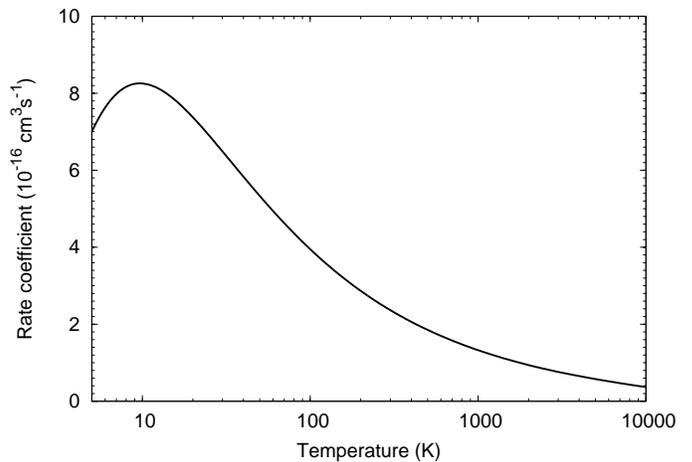}
  \caption{Rate constant for the radiative association process $b\ ^3\Sigma^+ \rightarrow a\ ^3\Sigma^+ + h\nu$.}
  \label{fig_rad_stab_rate_bstate}
\end{figure}

The rate coefficient, calculated using expression (\ref{eq_rad_stab_rate}), is shown in Fig. \ref{fig_rad_stab_rate_bstate}. It presents a maximum at 10~K and decreases at higher temperature. It is smaller than the rate coefficient for association in the ground state \cite{Zygelman1990,Kraemer1995}. However, the value of the rate is probably slightly larger than the one presented in this work due to the contribution of shape resonances that we neglected.

\begin{figure}
\includegraphics[angle=-90,width=.5\textwidth]{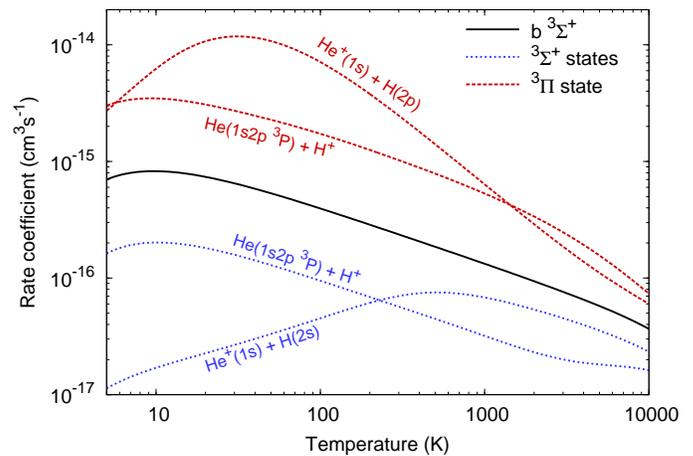}
  \caption{Rate constants for radiative association of HeH$^+$ in the $ a\ ^3\Sigma^+$ state along various electronic states. Black full line: association along the $b\ ^3\Sigma^+$ state; dotted blue line: association along other $^3\Sigma^+$ state; dashed red line: association along $^3\Pi$ states.}
  \label{fig_rad_stab_rate}
\end{figure}

As we have calculated the partial photodissociation cross sections into all the $n=1-3$ excited states, we can easily evaluate the cross section for radiative association in a discrete level of the $a\ ^3\Sigma^+$ state following a spontaneous transition from the continuum of  any of the excited electronic states. In Fig. \ref{fig_rad_stab_rate}, we show the rate coefficients for the dominant channels of formation of HeH$^+$ in the $a\ ^3\Sigma^+$ state. It should be noted that it is necessary to include the effect of the non-adiabatic couplings to obtain accurate rate coefficients, due to the strong interaction between the excited electronic states of HeH$^+$.
In the $^3\Sigma^+$ symmetry, the radiative association process is dominated by formation along the $b\ ^3\Sigma^+$ state. The electronic channels that give the next largest rate coefficients for the radiative association of the $a\ ^3\Sigma^+$ state are the second and fourth $n=2$ states (see Fig. \ref{fig_elec_states}), which correlate asymptotically to He($1s2p\, ^3P^{\text{o}}$) + H$^+$ and He$^+(1s)$ + H($2s$), respectively. However, the rates are much smaller for these states.
On the other hand, the approach along the two lowest $^3\Pi$ states (dissociating into He($1s2p\, ^3P^{\text{o}}$) + H$^+$ and He$^+(1s)$ + H($2p$), respectively) yields rate coefficients that are much larger than the rate corresponding to the transition $b\ ^3\Sigma^+ \rightarrow a\ ^3\Sigma^+$. 
This shows that radiative association along the $b\ ^3\Sigma^+$ is not the most efficient way of producing HeH$^+$ in the $a\ ^3\Sigma^+$ state.

\subsection{Applications}

Metastable helium can be present in significant concentration in astrophysical environments such as planetary nebulae due to recombination of He$^+$ with electrons \cite{Osterbrock2006}, and it is conceivable that it might influence the abundances of various atomic and molecular species. Roberge and Dalgarno \cite{Roberge1982} already considered the role of metastable helium on the production of HeH$^+$ in the ground state following collisions between He($1s2s\ ^3S$) and atomic or molecular hydrogen and showed that under some circumstances, these reactions can be a substantial source of HeH$^+$.

Based on the rates presented in Sec. \ref{section_results}, we can provide a rough estimate of the abundance of HeH$^+(a\ ^3\Sigma^+)$ in planetary nebulae. If we assume equilibrium between photodissociation and radiative association along the $b\ ^3\Sigma^+$ state, we have
\begin{equation}
n(\text{HeH}^+) k^{\text{d}} = n(2\, ^3S) n(\text{H}^+) k^{\text{a}}
\end{equation}
where $n(\text{HeH}^+)$, $n(2\, ^3S)$, and $n(\text{H}^+)$ are the density of HeH$^+$, He($1s2s\, ^3S)$ and H$^+$, respectively, while $k^{\text{d}}$ and $k^{\text{a}}$ are the rates for the photodissociation and radiative association processes.
We consider here a typical nebula with parameters $n(\text{H}^+)=10^4$ cm$^{-3}$, $n(\text{He}^+)=10^3$ cm$^{-3}$, $T_\star =5\times10^4$~K and $T_e=10^4$~K \cite{Drake1972}.
A formula giving the abundance of metastable helium in nebulae as a function of the He$^+$ density and electron density and temperature was derived by Clegg \cite{Clegg1987}. Using the parameters above, we get a density of metastable helium $n(2\, ^3S)\approx 4\times 10^{-3}$ cm$^{-3}$. Combining with the results for the rates $k^{\text{d}}$ and $k^{\text{a}}$ given respectively in Tab. \ref{table_photo_rate} and Fig. \ref{fig_rad_stab_rate_bstate}, we get $n(\text{HeH}^+)=3\times 10^{-10}$ cm$^{-3}$. The density of HeH$^+$ in its triplet state is much smaller than in the ground state \cite{Cecchi-Pestellini1993} but it can nonetheless influence the abundance of other species through reactions (\ref{eq_photodissociation}) and (\ref{eq_radiative_association}).
The radiative association process (\ref{eq_radiative_association}) could also be of importance in the chemistry of the early universe \cite{Lepp2002}. Neutral helium was first produced following recombination of He$^+$, therefore populating the $2\, ^3S$ state. While metastable helium was probably re-ionized due to its low binding energy, it could still form HeH$^+$ by radiative association with H$^+$.

\section{Conclusions}
We have investigated the photodissociation of HeH$^+$ in its metastable triplet state by means of time-dependent methods. We used previously reported potential energy curves, non-adiabatic couplings and dipole moments of the molecular ion HeH$^+$ to calculate the cross section and rate coefficient for this process using a wave packet approach. We found that the photodissociation cross sections and rate coefficients are dominated by the contribution of the excited states, and in particular the $^3\Pi$ states. 
The cross section and rate coefficients for the inverse process, radiative association, were computed on the basis of the photodissociation cross sections. HeH$^+$ can be formed in its triplet state by association of metastable helium and a proton, a process which is likely to occur in various astrophysical environments. Based on these results, we estimated the abundance of this triplet state and found it to be much lower than HeH$^+$ in its ground state.

\acknowledgments
This work was supported by the U.S. Department of Energy and by the Communaut\'e fran\c caise of Belgium (Action de Recherche Concert\'ee) and the Belgian National Fund for Scientific Research (FRFC/IISN Convention). 
S. Vranckx would like to thank the FRIA-FNRS for financial support.

\end{document}